# Novel Lithium-Sulfur Polymer Battery Operating at Moderate Temperature

Vittorio Marangon[a], Daniele Di Lecce[b], Luca Minnetti[b], Jusef Hassoun[a,b,c]*

[a] *Department of Chemical, Pharmaceutical and Agricultural Sciences, University of Ferrara, Via Fossato di Mortara 17, Ferrara, 44121, Italy*

[b] *Graphene Labs, Istituto Italiano di Tecnologia, via Morego 30, Genova, 16163, Italy*

[c] *National Interuniversity Consortium of Materials Science and Technology (INSTM), University of Ferrara Research Unit, Via Fossato di Mortara, 17, 44121, Ferrara, Italy.*

*Corresponding author. E-mail addresses: jusef.hassoun@unife.it, jusef.hassoun@iit.it.

**Abstract**

A safe lithium-sulfur battery employs a composite polymer electrolyte based on a poly(ethylene glycol) dimethyl ether (PEGDME) solid at the room temperature. The electrolyte membrane enables a stable and reversible Li-S electrochemical process already at 50 °C, with low resistance at the electrode/electrolyte interphase and fast Li$^+$ transport. The relatively low molecular weight of the PEGDME and the optimal membrane composition in terms of salts and ceramic allow a *liquid-like* Li-S conversion reaction by heating at moderately high temperature, still holding the *solid-like* polymer state of the cell. Therefore, the electrochemical reaction of the polymer Li-S cell is characterized by the typical dissolution of lithium polysulfides into the electrolyte medium during discharge and the subsequent deposition of sulfur at the electrode/electrolyte interphase during charge. On the other hand, the remarkable thermal stability of the composite polymer electrolyte (up to 300 °C) suggests a lithium-metal battery with safety content significantly higher than that using the common, flammable liquid solutions. Hence, the Li-S polymer battery delivers at 50 °C and 2 V a stable capacity approaching 700 mAh g$_S^{-1}$, with steady state coulombic efficiency of 98%. These results suggest a novel, alternative approach to achieve safe, high-energy batteries with solid polymer configuration.

**Table of Content**

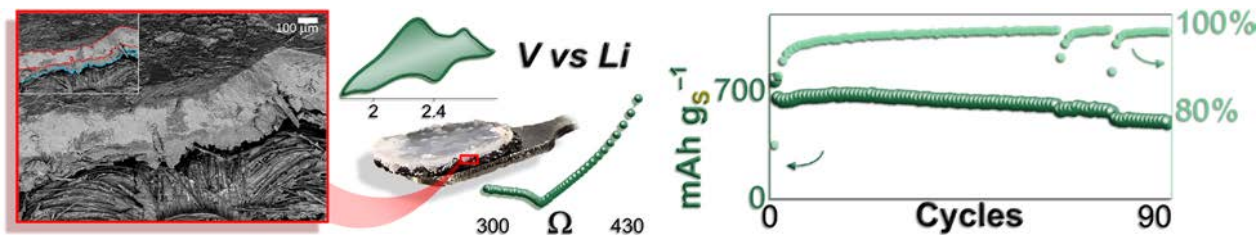

**A lithium-sulfur polymer battery** delivers at 50 °C almost 700 mAh $g_S^{-1}$ over 90 charge/discharge cycles. The polymer electrolyte has high thermal stability and forms a stable interphase with the electrodes suitable for the Li-S conversion process at 2.2 V *vs*. $Li^+/Li$. The new electrolyte formulation may enable safe, high-energy electrochemical cell.

**Keywords**

Poly(ethylene glycol) dimethyl ether; solid PEGDME; composite polymer electrolyte; Li-S battery; lithium polymer battery.

**Introduction**

The lithium-sulfur (Li-S) cell is one of the most promising next-generation battery systems, benefiting from the multi-electron conversion process $S_8 + 16Li^+ + 16e^- \rightleftarrows 8Li_2S$,[1,2] which leads to a theoretical energy density as high as 2600 Wh $kg^{-1}$, when referred to the $Li_2S$ mass.[3] This reaction occurs via formation of several lithium polysulfides ($Li_2S_x$, $2 \leq x \leq 8$)[4,5] which are soluble in the electrolyte solution for $x \geq 4$ and may migrate to the lithium anode, causing active material loss and cell degradation by reaction at the electrode surface.[6] Therefore, significant effort has been devoted over the past decade to optimize suitable cathode architectures that may mitigate the low electronic conductivity of S, $Li_2S_2$, and $Li_2S$ and provide suitable reaction sites for the lithium polysulfides.[7] So far, a large variety of electrode compositions involving various additives, such as carbon matrices,[8–12] MOFs,[13–15] and metal nanoparticles,[16,17] have been proposed. On the other hand, the addition of sacrificial agents to the electrolyte solution, e.g., $LiNO_3$,[18,19] has proven to be the key strategy to actually mitigate the detrimental effects of polysulfide dissolution via formation of a stable solid electrolyte interphase (SEI) on the lithium anode.[20] Furthermore, electrolyte additives may inhibit metallic-dendrite growth, that is, an undesired process hindering the large-scale application of rechargeable lithium-metal batteries.[20,21] Significant improvements have been achieved by developing alternative electrolyte formulations to those employed in conventional lithium-ion batteries, which react with the lithium polysulfides.[22] In this regard, liquid solutions of lithium salts with large anions and $LiNO_3$ in 1,3-dioxolane:1,2-dimethoxyethane (DOL:DME) mixtures are chemically stable towards lithium polysulfides, form a suitable film over the anode, have a low viscosity and high ion conductivity, thereby enabling high performance of the Li-S cell.[23] However, excessive volatility and high flammability of these interesting electrolyte media pose safety concerns, in particular in cells using lithium metal.[10] Alternative electrolyte systems employing liquid end-capped glymes as solvents, such as poly(ethylene glycol) dimethyl ether (PEGDME, $CH_3O(CH_2CH_2O)_nCH_3$) with a low *n* value,[24–26] have a reasonably high flash point.[27] On the other hand, long-ether-chain PEGDME with average molecular weight (MW) higher than 1000 g mol$^{-1}$ (moderately high *n* value) is a solid with semi-crystalline structure, low flammability, and negligible volatility at room temperature. Solid polymer electrolytes using poly(ethylene-oxide) (PEO, high *n* value) have remarkable thermal, mechanical, and electrochemical stability, along with high compatibility with various lithium salts and enhanced Li$^+$ ions transport.[28] However, excessive crystallinity at room temperature due to the relatively high molecular weight (that is, typically higher than 100000 g mol$^{-1}$) limited the large-scale application PEO-based polymer electrolytes, to date commercialized in niche automotive and stationary-storage sectors that allow a higher operating temperature.[21] Indeed, suitable ionic conductivity (> $10^{-4}$ S cm$^{-1}$) and adequate Li-ion transport properties are typically achieved at the predominantly amorphous state of the EO chains above 65 °C,[29] which allows battery application employing insertion[30–33] or sulfur-based conversion cathodes.[34–39] The use of ceramic fillers such as $SiO_2$, $Al_2O_3$, $TiO_2$ or $ZrO_2$ generally promotes the amorphous phase in PEO and enhances the ionic conductivity, although the polymer cannot be applied below 60 °C.[29,40,41] In this regard, solid PEGDME has mechanical and chemical stability compatible with lithium cell application, and a melting point allowing operation at lower temperature compared to PEO.[42] Indeed, in our earlier report a Li|LiFePO$_4$ battery exploited a PEGDME-based composite polymer electrolyte (PEGDME_CPE) incorporating ceramic SiO$_2$ particles, lithium bis(trifluoromethanesulfonyl)imide (LiTFSI) as conductive salt, and LiNO$_3$ as SEI-forming agent.[42] The above cell operated at 50 °C and delivered reversibly 125 mAh g$^{-1}$, with a capacity retention as high as 99% over 300 cycles.[42] This electrolyte formulation benefitted from advantageous electrochemical and thermal properties that set between those of the well-known solid PEO-based electrolytes (PEO_PEs)[41] used in the typical Li-polymer cell, and the common DOL:DME-based liquid electrolytes (DOL:DME_LEs) employed in the Li-S battery.[23] Notably, the former electrolytes may significantly enhance the safety content of the cell with respect to the latter, although they typically suffer from poor Li$^+$ transport properties at moderate temperatures.[41] Therefore, we explore in this work the applicability of the PEGDME_CPE in high energy Li-S batteries working at moderate temperature (i.e., 50 °C). It is worth mentioning that a Li-S cell exploiting a solid polymer electrolyte synthesized by using a liquid PEGDME with average MW of 250 g mol$^{-1}$ as the starting precursor was previously achieved.[43] On the other hand, the Li-S cell reported herein originally employs a pristine PEGDME with average MW of 2000 g mol$^{-1}$ as solid polymer matrix to achieve the composite membrane.[44] We investigate the ion transport, the thermal properties and the electrode/electrolyte interphase, and demonstrate the efficient operation at 50 °C of the lithium-sulfur polymer battery which is allowed by the relatively low molecular weight of the selected PEGDME.

**Results and discussion**

To better evaluate the differences between the various electrolyte configurations, Figure 1 reports a comparison between solid PEO_PE[41] and PEGDME_CPE, and a DOL:DME_LE solution in terms of Li$^+$ transference number (t$^+$) measured at the optimal operative temperature of each medium (Fig. 1a) and ionic conductivity at 50 °C (Fig. 1b). In this work, the "optimal operative temperature" refers to the most adequate temperature value for allowing efficient operation in a Li-S cell depending on the employed solvent, that is, 25 °C for DOL:DME solutions,[23,45–47] 50 °C for the PEGDME2000,[42] and 80 °C for PEO.[41] The DOL:DME_LE exhibits a significantly higher $t^+$ at room temperature compared to that of PEGDME_CPE at 50 °C, that is, 0.67 *vs.* 0.23 (Fig. 1a and Figure S1 in Supporting Information),[42] while the PEO_PE has a $t^+$ of 0.22 at 80 °C.[41] Relevantly, the PEGDME_CPE has ionic conductivity exceeding by two order of magnitude that of PEO_PE at 50 °C ($1 \times 10^{-4}$ with respect to $2 \times 10^{-6}$ S cm$^{-1}$ in Fig. 1b).[41,42] Despite being adequate for battery application, the conductivity of PEGDME_CPE at the latter temperature is 10 times lower than that of the DOL:DME_LE at 25 °C due to the higher mobility of the ions in the solution compared to the solid polymer (see Fig. S1b in Supporting Information for further details on these measurements).[42] These data suggest that the PEGDME_CPE might enable satisfactory ion motion at moderate temperature, which is considered a key requirement for application in lithium-sulfur polymer cells,[21] although possible issues in the experimental determination of $t^+$ leading to an overestimation of the calculated value cannot be completely excluded.[48,49] We remark that the relatively low molecular weight of the PEGDME used for achieving the polymer electrolyte (average MW = 2000 g mol$^{-1}$, see the Experimental section for further details) may actually allow suitable battery performance, superior safety level, and a lithium-sulfur conversion process below the conventional application temperature of PEO-based electrolytes.[50] Indeed, PEO_PE requires a higher temperature than PEGDME_CPE to reach a comparable $t^+$ (i.e., 80 °C rather than 50 °C in Fig. 1a) due to the relevantly higher molecular weight (i.e., 600000 g mol$^{-1}$ *vs* 2000 g mol$^{-1}$). Furthermore, Figure S2 in the Supporting Information also demonstrates the applicability of the PEGDME_CPE at higher temperatures, as the electrolyte exhibits an ionic conductivity of $4 \times 10^{-4}$ S cm$^{-1}$ at 80 °C. In this regard, the PEGDME_CPE shows high thermal stability, as confirmed by thermogravimetric analysis (TGA) under inert atmosphere (see Fig. 1c and Fig. 1d showing the corresponding differential thermogravimetry, i.e., DTG). The thermal analysis of the solid PEGDME2000 polymer, reported in the same figure for comparison, reveals a single weight loss starting at about 280 °C and centered slightly above 380 °C, which suggests a large application temperature range. On the other hand, the PEGDME_CPE exhibits three weight-loss steps: the first one is centered at 340 °C and might be ascribed to removal of the fraction of PEGDME2000 chains that interact with the $SiO_2$ particles;[51] the second one occurs at 400 °C and is likely related to complexes formed between PEGDME2000 and the solvated lithium salts, which are characterized by higher decomposition temperature;[52] the third one is observed in the temperature range from 420 °C to 470 °C and is mainly attributed to the decomposition of LiTFSI.[53] Notably, 15 wt.% sample residues after the measurement mostly consist of $SiO_2$ along with chemical compounds formed by the decomposition of the lithium salts (LiTFSI and $LiNO_3$).[53] Therefore, our composite polymer electrolyte ensures a high ionic conductivity, a suitable cation transference number for application in batteries exploiting the lithium metal anode, and a remarkable thermal stability (up to 280 °C).

**Figure 1**

The electrochemical response of the Li-S polymer battery employing the PEGDME_CPE is herein investigated by cyclic voltammetry (CV) and electrochemical impedance spectroscopy (EIS) from 50 to 80 °C, in order to demonstrate the actual applicability and stability of this cell configuration in a wide temperature range. Figure 2 (a, c and e) reports the voltammograms of the cell tested consecutively at 50, 60 and 70 °C, respectively, whilst Figure S2a in Supporting Information shows that measured at 80 °C. Figure 1a reveals for the polymer battery a discharge process at 50 °C with current peaks at 1.94 and 2.33 V *vs.* $Li^+$/Li upon the first cycle, which are reflected as two charge signals at 2.28 and 2.60 V *vs.* $Li^+$/Li. This response is in part consistent with that of conventional Li-S batteries using liquid electrolyte solutions (Fig. S4a in the Supporting Information), typically characterized by two discharge peaks at about 2.3 and 2.0 V *vs.* $Li^+$/Li, reflecting the reversible conversion of lithium and sulfur to soluble lithium polysulfides with various chain lengths ($Li_2S_x$ with $4 \leq x \leq 8$) along with solid $Li_2S_2$ and $Li_2S$, and two overlapped charge signals around 2.3 V *vs.* $Li^+$/Li.[54] The subsequent CV profiles of the polymer battery at 50 °C (Fig. 2a) exhibit a shift of the discharge peaks to higher potentials likely due to improved reaction kinetics, and a raise of the charge peak at 2.28 V *vs.* $Li^+$/Li suggesting an enhancement of the polysulfide oxidation rate. According to earlier reports on Li-S cells using electrolyte solutions,[16,17,55,56] this activation upon cycling can be associated with micro-structural rearrangements in cathode which increase the sulfur utilization and enhance the electrode/electrolyte interphase. Further improvements in electrochemical activity can be achieved by rising the operating temperature, as shown in Figure 2c and 2e, as well as in Figure S3a of the Supporting Information. Indeed, an increase in temperature up to 60 (Fig. 2c), 70 (Fig. 2e) and 80 °C (Fig. S3a) leads to a higher peak current upon charge and to a better overlap of the CV profiles (in particular at 60 and 70 °C), thus suggesting a very stable and reversible electrochemical process. Furthermore, a discharge peak appears slightly above 2 V *vs.* Li$^+$/Li when the temperature ranges from 60 to 80 °C, thereby indicating that the segmental motion of the polymer chains may assist the ionic transport through the electrolyte and at the electrode/electrolyte interphase during the conversion process.[21] Furthermore, the presence of an additional reduction peak between 2.0 and 2.1 V *vs.* Li$^+$/Li observed in the CV measurements performed at 60, 70 and 80 °C, whose profiles are reported in Figures 2c, 2d and S3a, respectively, as well as the absence of a zero-current response current at the end of cathodic scan, is likely ascribed to a relatively sluggish kinetics for the formation of the various polysulfides in the PEGDME_CPE matrix, and actually indicates a not fully reversible reaction. On the other hand, the reduction of S$_8$ to polysulfides in the lithium cell occurs through various reaction steps and leads to Li$_2$S$_x$ species with different chain length within the full range of the discharge potential. These steps may be partially detected by voltammetry depending on the operating temperature, as indeed shown in Figure 2. In particular, the additional peaks at the lower potential during reduction may be reasonably ascribed to the kinetically hindered formation of low-chain polysulfides (such as Li$_2$S$_4$ and Li$_2$S$_2$). Notably, the significant polarization of the polymer cell upon charging at 50 °C leads to two distinct charge peaks in the CV (Fig. 2a), while that employing the DOL:DME_LE (Figure S4a in the Supporting Information) displays broad, convoluted peaks during oxidation, which is in full agreement with the literature.[57] Therefore, the different electrochemical response of the Li-S polymer cell compared to the Li-S conventional one using a liquid solution may be in part ascribed to the mobility of Li$^+$ ions within the electrolyte medium and at the interphase between electrolyte and sulfur electrode. Accordingly, the polymer battery exhibits an electrochemical response approaching the *liquid-like* behavior as the temperature is gradually increased up to 80 °C. In this regard, further insight is given by Figure S5 and Table S4 in Supporting Information, which show that the above discussed voltammetry peaks shift toward each other at elevated temperature, whilst the discharge peaks mostly attributed to the long chain polysulfides shift at high potential. In particular, the 1$^{st}$ charge peak moves from 2.29 V *vs.* Li$^+$/Li at 50 °C to 2.35 V *vs.* Li$^+$/Li at 80 °C (Fig. S5b), while the 2$^{nd}$ charge signal from 2.60 to 2.52 V *vs.* Li$^+$/Li (Fig. S5c), thereby leading to a gradual merging of these peaks promoted by the increase in PEGDME chains mobility and Li$^+$ ions conductivity.[42] We remark that the polymer cell at 80 °C exhibits in CV a *liquid-like* performance (Fig. S3a) similar to that of Li-S batteries employing low-molecular-weight glymes, typically characterized by a higher viscosity than that of conventional DOL:DME mixtures.[58] Remarkably, the tests reported in Fig. 2a, c, e, and in Fig. S3a (Supporting Information) refer to 12 consecutive CV runs at various temperatures, thus suggesting a notable reversibility of the electrochemical process.

EIS measurements performed upon CV (see the Nyquist plots in Fig. 2b, 2d, 2f, and Fig. S3b the Supporting Information) reveal modifications at the electrode/electrolyte interphase in the Li-S polymer cell during cycling. Table 1 reports the results of NLLS analyses of the corresponding spectra, in terms of equivalent circuits (i.e., R$_e$(R$_i$Q$_i$)Q$_w$, see Experimental section), resistance values (i.e., R$_e$ and R$_i$), and $\chi^2$ parameter. The overall electrode/electrolyte interphase resistance (R$_1$ + R$_2$) of the polymer cell at 50 °C is represented by the width of the high-middle frequency semicircles of the Nyquist plot (Fig. 2a), and incorporates contributes by SEI film, electrode charge transfer, and possible grain boundaries.[59] The above resistance drops from 83 Ω at the open circuit voltage (OCV) condition to stable values around 42 Ω after 3 voltammetry cycles (Table 1). This behavior indicates the occurrence of activation processes and consequent enhancement of the electrode/electrolyte interphase by favorable microstructural modifications upon cycling.[16,17,55,56] The rise in operative temperature leads to a further decrease in the overall electrode/electrolyte resistance, thus reflecting the above-described improvements in conversion kinetics[42] (see Fig. 2d, 2f and S3b in the Supporting Information), with final values of 17, 14 and 12 Ω at 60, 70, and 80 °C (see Tables 1 and S2 in the Supporting Information). In this regard, the resistance trends at various temperatures reported in Figure S6 in Supporting Information show that the charge transport through both the interphase and the electrolyte (panels a and b, respectively) are thermally activated processes. Indeed, the electrolyte interphase resistance measured after subsequent voltammetry cycles decreases from 304 Ω at 50 °C to 150 Ω at 60 °C, 90 Ω at 70 °C, and 55 Ω at 80 °C. The thermal activation of the charge transfer process is also observed by the increase of the ($R_iQ_i$) elements in the equivalent circuit (Table 1 and Table S2 in the Supporting Information). Indeed, the difference between the circuits at 50 °C (Fig. 2b) and at higher temperatures (Fig. 2d, f, and Fig. S3b in Supporting Information) reflects the modifications of the electrode/electrolyte interphase achieved through temperature rise, which cause the deconvolution of the high-middle frequency semicircle into various contributes (SEI film, electrode charge transfer, and possible grain boundaries) represented by additional ($R_iQ_i$) elements. Relevantly, these changes are in line with the modification of the CV characteristics of the corresponding Li-S cell (Fig. 2a, c, e and Fig S3a in Supporting Information). In addition, both the benchmark Li-S cell at room temperature using DOL:DME_LE and the polymer Li-S cell at 80° C using PEGDME_CPE exhibit comparably low electrode/electrolyte interphase resistance, i.e., 2 Ω for the former and 12 Ω for the latter, which confirms that the polymer battery approaches the *liquid-like* condition. However, the ion motion is likely much faster in the benchmark electrolyte which shows related resistance of 5 Ω, with respect to the polymer one which has a resistance of 55 Ω (see Tables S2 and S3 in the Supporting Information). It is worth mentioning that the contribution of the lithium anode in the EIS measurements is not negligible. Indeed, the discussions of Figures 2, S3 and S4 in the Supporting Information refer to general electrode/electrolyte interphase to include both the cathode and the lithium side, whilst the study of the interphase formed between lithium and the PEGDME_CPE was already deeply considered in our late work,[42] where EIS and lithium stripping/deposition tests performed on symmetrical Li|PEGDME_CPE|Li cells at 50 °C revealed a

stable electrode/electrolyte interphase with low resistance and overvoltage. The affinity between lithium metal and the PEGDME_CPE is well confirmed by the EIS measurements reported in Fig. 2 in which all the Li-S cells exhibit low overall resistance values and stable kinetics. Therefore, both the CV and the EIS data reveal that the electrochemical activity of the Li-S polymer battery is adequate at 50 °C, and enhanced by thermal activation within the wide operative temperature range extending from 50 to 80 °C.

**Figure 2**

**Table 1**

Figure 3 illustrates the characteristics of the electrode/polymer-electrolyte assembly by showing photographic images of a sample before and after cycling (panel a and b), along with TGA data (panel c and d), respectively, and SEM images (panel e; see the Experimental section for details on sample preparation). A direct comparison between the PEGDME_CPE host on the positive electrode in the pristine condition and after consecutive CV cycles performed at 50, 60, 70, and 80 °C reveals a change in color of the polymer membrane from white (Fig. 3a) to a dark red (Fig. 3b), which demonstrates the expected dissolution of the lithium polysulfides into the electrolyte upon the electrochemical process.[60] It is worth mentioning that the dissolution of lithium polysulfides into the electrolyte, which is common process and not fully avoidable, does not necessarily lead to a compromising shuttle effect. In fact, the shuttle process that usually leads to an unlimited anodic reaction without charge accumulation and efficiency decrease is not observed in our tests, as indeed expected by the inclusion of the $LiNO_3$ additive into the PEGDME_CPE formulation, which promotes the formation of a protective SEI layer on the lithium surface.[25] Relevantly, literature work reported a high performance semi-liquid Li-S battery exploiting lithium polysulfides directly dissolved into the electrolyte without any shuttle reaction, due to the above-mentioned protection of the lithium surface by $LiNO_3$ additive.[58] TGA of the pristine and cycled PEGDME_CPE/electrode samples (Fig. 3c and 3d) indicates various weight loss processes, which are attributed to the components of the assembly. Thus, the pristine sample (black curves in Fig. 3c and 3d) shows a weight loss between 230 and 300 °C due to sulfur evaporation from the positive electrode,[61,62] along with two subsequent processes at 340 and 400 °C ascribed to the PEGDME2000 powder, which are in agreement with the TGA of Figure 1. Furthermore, weight variations ascribable to LiTFSI are observed between 420 and 445 °C,[53] and partial degradation of the electrode support likely occurs at about 560 °C. A residue of 27% of the initial weight after the heating scan accounts for $LiNO_3$,[53] $SiO_2$ particles, and electrode support. The cycled sample exhibits rather different thermal behavior (orange curves in Fig. 3c and 3d) characterized by an increase in weight at about 130 °C, which likely reflects reactions between $N_2$ and the lithium polysulfides during the thermogravimetric experiment, followed by a slight decrease due to sulfur evaporation up to 300 °C.[61,62] In addition, the loss of PEGDME2000 and LiTFSI appears as a single process centered at 390 °C, rather than the multiple losses between 340 and 445 °C observed for the pristine sample, thus suggesting changes in electrolyte composition by cycling in line with the macroscopic modifications displayed in Figure 3a and 3b. A residual mass of 43% is measured after complete degradation of the electrode support at 560 °C, accounting perhaps for possible crystalline $Li_2S$ (melting point = 940 °C). The modifications of the cathode/polymer interface during cycling in the cell are further investigated by SEM in Figure 3e, which shows at the left-hand side a cross section image of a sample recovered after consecutive CV between 50 and 80 °C and on the right-hand side panel a graphic representation of the cathode/polymer-electrolyte stack with indication of investigated area. In agreement with earlier reports,[16,17,55] the SEM image of the composite clearly shows various layers, which are attributed to (i) the fibrous carbon-cloth electrode support (below the cyan mark in Fig. 3e inset), (ii) the carbon coating of the electrode support along with the sulfur-carbon cathode film (gray layer between the cyan and the red marks in Fig. 3e inset), and (iii) a portion of the PEGDME_CPE membrane (light gray layer between red marks in Fig. 3e inset). Notably, our electron microscopy data suggest improved contact between the PEGDME_CPE and the sulfur-carbon electrode, which may favor the charge transfer at the interphase, as well contact regions in the electrolyte layer (colored by dark gray) which might be associated with the dissolution of lithium polysulfides during the electrochemical process. It is worth mentioning that PEGDME_CPE was not involved in the electrode formulation and the only polymeric species included in the cathode formulation is poly(vinylidene fluoride) (PVDF), which acts as binding agent (see experimental section), although the addition of a polymer electrolyte to the cathode composition is a well-known technique to achieve the formation of an enhanced electrode/electrolyte interphase.[43] Therefore, the formation of a suitable electrode/electrolyte interphase may be promoted by the relevant mobility of the PEGDME polymer chains at 50 °C, which is sufficient for allowing a proper cathode wetting and $Li^+$ ions exchange. Furthermore, a contribution to the $Li^+$ ions conductivity given by blending between PEGDME and PVDF cannot be excluded. On the other hand, the formation of a stable electrode/electrolyte assembly is suggested by the SEM image in Figure 3e, which displays an adequate contact between the electrolyte and the cathode film.

**Figure 3**

The galvanostatic-cycling performance of the Li-S polymer battery at 50 °C is herein evaluated at the current rate of C/10 (1C = 1675 mA $gs^{-1}$). These testing conditions represent an optimal choice, which might match the typical requirements of the stationary storage market. We remark that the polymer configuration would ensure enhanced thermal stability, possibly allowing a safe use in large battery packs, whilst low current rate and moderately high temperature are well compatible with load-balancing applications in smart grids.[21] Figure 4 shows the voltage profiles (panel a) and cycling behavior (panel b) of the above Li-S polymer cell, which steadily delivers a satisfactory capacity with high coulombic efficiency. In more detail, Fig. 4a reveals the partial merging of the two characteristic plateaus at about 2.4 and 1.8 V upon the first discharge, accounting for the conversion of lithium and sulfur to lithium polysulfides (see CV in Fig. 2a),[63] as well as two definite charge plateaus at about 2.3 and 2.6 V. The sloping shape of the discharge plateau may suggest moderate $Li^+$ diffusion hindering within the PEGDME_CPE as well as slow stabilization of the electrode/electrolyte interphase,[64] which is gradually improved upon the subsequent cycles. Indeed, microstructural reorganizations due to polysulfide dissolution upon cycling[16,17,55,56] favors a gradual change in cell response, leading to well-defined discharge plateaus at 2.4 and 1.9 V (Fig. 4a).

Furthermore, subsequent overlapping voltage curves characterized by moderate polarization between charge and discharge indicate a stable and efficient electrochemical process. In this regard, Figure 4b reveals a maximum capacity of 770 mAh $g_s^{-1}$, with a steady-state value ranging between 700 and 600 mAh $g_s^{-1}$, and a retention of 71% for over 90 cycles. Interestingly, the cell shows after the initial activation cycles a coulombic efficiency approaching 99%, except for few intermediate cycles particularly occurring at the final stages of the test, characterized by values decreasing down to about 90%, and raising again to the steady state (Fig. 4b, light-green curve and right-side axis). This decrease in efficiency likely suggests the occurrence of micro-dendrites which are efficiently suppressed by the polymer electrolyte with the ongoing of the cycling to achieve the pristine efficiency value.[42] This important aspect represent an additional bonus, in particular concerning the high safety content, of the PEGDME_CPE proposed herein for Li-S battery application. The stability of the electrode/electrolyte interphase is further demonstrated in Fig. 4 (c and d), which reports the Nyquist plots at 50 °C of a Li-S cell employing the PEGDME_CPE at the OCV condition and after 50 cycles at C/10 (1C = 1675 mA $g_s^{-1}$), respectively, while the related results of a NLLS analysis are displayed in Table 2. These Nyquist plots reveal a drop of the interphase resistance ($R_1$) upon cycling from about 180 Ω to 110 Ω (high-frequencies grain boundaries due to partial crystalline phase were not considered), denoting enhancements of the electrochemical activity at the electrode/electrolyte interphase by cycling in agreement with the EIS data reported in Figures 2 and S3. In addition, the portion of cycled PEGDME_CPE/electrode sample recovered from the cell after 50 cycles is shown in Fig. 4e, which reveals the formation of a blend between the electrolyte and the electrode evidencing suitable contact, while the dark red color of the cycled PEGDME_CPE membrane confirms the uniform dissolution of the lithium polysulfides during cell activity without any shuttle process, as indeed observed in Fig. 3. These data clearly demonstrate the suitability of the Li-S system and display the stability of the electrode/electrolyte interphase upon cycling. Thus, considering an average capacity of 600 mAh $g_s^{-1}$ and an electrochemical process centered at 2.2 V, the Li-S polymer cell has a theoretical energy density of about 1300 Wh $kg^{-1}$, which might lead to a suitable practical energy density and high efficiency for applications that require high-thermal stability.[21] On the other hand, challenging operative conditions including high temperature may hinder the application of typical Li-S battery configurations based on volatile liquid electrolytes, as demonstrated in Figure S7 in the Supporting Information which shows a discharge/cycling test at the constant rate of C/10 (1C = 1675 mA $g_S^{-1}$) performed at 50 °C on a Li-S battery employing the DOL:DME_LE. The voltage profiles displayed in Fig. S7a reveal an irreversible discharge step at about 1.9 V likely ascribed to the reduction of $LiNO_3$ dissolved in the electrolyte solution promoted by the relatively high temperature, since this plateau it is usually observed around 1.6 V at room temperature.[58] The excessive passivation of the lithium surface by reduction of $LiNO_3$ may cause the formation of a thick electrode/electrolyte interphase, as suggested by the decay of capacity to around 900 mAh $g_S^{-1}$ (Fig. S7b), which is a lower value with respect to the one delivered by the Li-S at 25 °C.[55] Furthermore, both DOL and DME solvents suffer from marked volatility which may be promoted by the challenging temperature value, whilst the best performance of this liquid electrolyte is achieved at 25 °C.[45–47,55] These data suggest the poor applicability of the DOL:DME solutions at high temperature, which, on the other hand, improves the performance of the PEGDME_CPE electrolyte.

**Figure 4**

**Conclusions**

To the best of our knowledge, we demonstrated for the first time that composite polymer electrolytes using PEGDME with MW of 2000 g $mol^{-1}$ as the solid solvent may be effectually applied in a Li-S battery operating at a temperature as low as 50 °C. The composite electrolyte membrane exhibited at 50 °C a $Li^+$ transference number of 0.23 and an ionic conductivity of $1\times10^{-4}$ S $cm^{-1}$, while the PEO-based benchmark electrolyte displayed similar $t^+$ values (0.22) at 80 °C and considerably lower ionic conductivity ($2\times10^{-6}$ S $cm^{-1}$) at 50 °C.[41] TGA evidenced a thermal stability extended up to 300 °C, suggesting suitable characteristics for applications requiring high safety level, such as the stationary energy-storage. Our data indicated that the polymer electrolyte forms a favorable interphase on both anode and cathode, which leads to a stable Li-S conversion process with low charge transfer resistance within the temperature range from 50 to 80 °C. In this interval, the polymer battery operated by electrochemical processes mainly centered at 2.4 and 2.0 V *vs.* Li$^+$/Li during discharge and at 2.3 and 2.6 V *vs.* Li$^+$/Li during charge, as revealed by CV, although above 60 °C we observed enhanced conversion kinetics, leading to better overlapping of the potential profiles and more intense current signals. This promising cell response was attributed to the suitably low resistance of both electrode/electrolyte interphase (between 83 and 12 Ω) and electrolyte (between 304 and 55 Ω), which was measured by EIS in the 50 – 80 °C range. Furthermore, increase in temperature gave rise to a third discharge step at 2.3 V *vs.* Li$^+$/Li along with a gradual shift of the charge peaks up to formation of a broad double-signal similar to that observed in Li-S cells using a conventional liquid electrolyte. Indeed, the polymer system revealed a profile change from *solid-like* to *liquid-like* Li-S battery, upon increasing the temperature from 50 °C to 80 °C. Accordingly, this work has provided evidence of lithium polysulfide dissolution into the electrolyte upon cell operation, which influenced the thermal and morphological characteristics of the cathode/electrolyte-membrane array. The Li-S polymer battery operated at 50 °C with a working voltage of 2.2 V, delivering a capacity above 600 mAh g$_S^{-1}$ at C/10 (1C = 1675 mA g$_S^{-1}$), with a retention of 71% for more than 90 discharge/charge cycles and a maximum coulombic efficiency of 98%. It is worth mentioning that the reversibility of the cell may be further improved by carefully tuning the electrolyte composition in terms of amounts of sacrificial additive (e.g., LiNO$_3$) and ceramic (e.g., SiO$_2$), in order to enhance the SEI layer on the electrodes surface. Therefore, our study suggests a new pathway to achieve safe lithium-metal cell exploiting the high-energy Li-S conversion process.

**Experimental**

*Achievement of the composite polymer electrolyte*

The composite polymer electrolyte (CPE) was prepared by mixing polyethylene glycol dimethyl ether (PEGDME2000, CH$_3$O(C$_2$H$_4$O)$_n$CH$_3$, average MW of 2000 g mol$^{-1}$, Sigma-Aldrich), lithium

bis(trifluoromethanesulfonyl)imide (LiTFSI, 99.95% trace metals basis, Sigma-Aldrich), lithium nitrate (LiNO$_3$, 99.99% trace metals basis, Sigma-Aldrich), and fumed silica (SiO$_2$, average particle size: 0.007 μm, Sigma-Aldrich). Before use, LiTFSI and LiNO$_3$ were dried under vacuum for 2 days at 110 °C. Either salts were added to the PEGDME2000 in the 1 mol kg$^{-1}$ concentration as referred to the mass of the latter, and 10 wt.% SiO$_2$ (as referred to the mass of PEGDME2000-lithium salts mixture) was incorporated in the CPE. The solid electrolyte membrane was obtained by forming a dense, semi-liquid slurry of the above components with acetonitrile (ACN, Sigma-Aldrich), which was subsequently removed upon several drying steps at various temperatures as reported in a previous work.[42] This CPE is herein indicated as PEGDME_CPE.

*Preparation of control liquid electrolyte*

A control liquid electrolyte (LE) was synthetized by dissolving LiTFSI (1 mol kg$_{solvent}^{-1}$) and LiNO$_3$ (1 mol kg$_{solvent}^{-1}$) in a solution of 1,3-dioxolane (DOL, anhydrous, containing ca. 75 ppm of butylated hydroxytoluene, i.e., BHT, as inhibitor, 99.8%, Sigma-Aldrich) and 1,2-dimethoxyethane (DME, anhydrous, inhibitor-free, 99.5%, Sigma-Aldrich) in the 1:1 weight ratio. Prior to use, both DOL and DME were dried by molecular sieves (3 Å, rod, size 1/16 in., Honeywell Fluka) until a water content below 10 ppm was obtained as measured by 899 Karl Fischer Coulometer, Metrohm, whilst LiTFSI and LiNO$_3$ were dried under vacuum for 2 days at 110 °C as above mentioned. This LE is herein indicated as DOL:DME_LE.

*Preparation of sulfur electrode*

The sulfur-carbon composite was prepared as described previously,[55] by mixing elemental sulfur (S, ≥99.5%, Riedel-de Haën) and conductive carbon black (Super P, Timcal, SPC) the weight ratio of 70:30 under magnetic stirring in a silicone oil bath at 120 °C. The resulting composite was subsequently cooled down to room temperature and ground in an agate mortar to obtain a fine powder. This composite is herein referred as S:SPC 70:30 w/w. Sulfur electrode disks were obtained through doctor blade casting (MTI Corp.) of a slurry formed by 80 wt% sulfur-carbon composite (i.e., S:SPC 70:30 w/w), 10 wt% conductive carbon black (Super P, Timcal, SPC), and 10 wt% poly(vinylidene fluoride) binder (Solef® 6020 PVDF) homogeneously dispersed in *N*-methyl-2-pyrrolidone (NMP, Sigma-Aldrich). The slurry was cast on a porous carbon-cloth foil (GDL ELAT 1400, MTI Corp.), which was then heated on a hot plate at 50 °C for about 3 h under a fume hood. Afterwards, electrode disks with diameter of 14 mm and 10 mm were cut out from the coated carbon-cloth and dried overnight at 35 °C under vacuum before being transferred in argon-filled glovebox (MBraun, H$_2$O and O$_2$ content below 1 ppm). The obtained sulfur loading on the electrodes was about 1 mg cm$^{-2}$.

*Cell assembly and electrochemical tests*

CR2032 coin-type cells (MTI Corp.) were assembled in an argon-filled glovebox (MBraun, H$_2$O and O$_2$ content below 1 ppm) and studied by using various electrochemical techniques. The electrolyte lithium transference number (t$^+$) was evaluated according to the Bruce-Vincent-Evans method,[65] by applying to a symmetrical Li|Li cell a voltage of 30 mV for 90 minutes (chronoamperometry measurement) and collecting impedance spectra of this cell before and after polarization. The alternate voltage bias of these EIS measurements had an amplitude of 10 mV, and the frequency investigated ranged from 500 kHz to 100 Hz. The t$^+$ value was calculated using equation (1):[65]

$$t^+ = \frac{i_{ss}}{i_0} \frac{\Delta V - i_0 R_0}{\Delta V - i_{ss} R_{ss}} \tag{1}$$

where ΔV is the chronoamperometry voltage (i.e., 30 mV), $i_0$ is the initial current during polarization (t = 0), $i_{ss}$ is the current after polarization (t = 90 min), and $R_0$ and $R_{ss}$ are the electrode/electrolyte interphase resistance determined by EIS before and after the 90-minute polarization, respectively. The ionic conductivity of the electrolytes was extracted from EIS data, which were collected by applying to a symmetrical cell with stainless steel (SS) electrodes an alternate voltage signal amplitude of 10 mV within the 500 kHz – 100 Hz frequency range.

Li|PEGDME_CPE|S:SPC 70:30 w/w cells were assembled by stacking a lithium disk with a diameter of 14 mm, with PEGDME_CPE membrane and S:SPC 70:30 w/w electrode having diameters of 10 mm housed into 4 polymeric O-rings (CS Hyde, 23-5FEP-2-50) with internal

diameter of 10 mm, and thickness of 127 μm each. Prior to the tests, all the Li|PEGDME_CPE|S:SPC 70:30 w/w cells were exposed to 4 heating-cooling cycles between 25 and 70 °C to decrease the crystallinity and enhance the ionic conductivity of the PEGDME_CPE; each cycle had a duration of 24 h (i.e., 12 h for each heating and cooling step).[42] Li|DOL:DME_LE|S:SPC 70:30 w/w control cells were assembled by employing a 14-mm diameter sulfur-carbon electrode separated from the lithium anode by a 16-mm diameter Celgard 2400 foil soaked with 25 μL of electrolyte solution for CV tests and 15 μL $mg_S^{-1}$ for galvanostatic cycling, as previously reported.[55] CV measurements were carried out on a Li|PEGDME_CPE|S:SPC 70:30 w/w cells at 50, 60, 70, and 80 °C, as well as on a Li|DOL:DME_LE|S:SPC 70:30 w/w control cell at room temperature, by employing a scan rate of 0.1 mV s$^{-1}$ in the 1.8 – 2.8 V vs. Li$^+$/Li potential range. EIS measurements were performed on these cells at the OCV condition as well as upon the voltammetry cycles at the above-mentioned temperature conditions, by applying an alternate voltage signal with amplitude of 10 mV within the 500 kHz – 100 mHz frequency range. The impedance spectra were analyzed with the Boukamp software using the non-linear least squares (NLLS) method (the $\chi^2$ was in the order of 10$^{-4}$ or lower).[66,67] The impedance response of the cell was modelled by using equivalent circuits which incorporate the high-frequency electrolyte resistance ($R_e$), high-to-middle frequency resistive and constant phase elements ($R_iQ_i$) arranged in parallel and ascribed to the electrode/electrolyte interphase, as well as low frequency resistive and capacitive elements accounting for the Warburg-type, Li-ion diffusion ($R_w$ and $Q_w$).[66,67] Galvanostatic cycling tests were performed on Li|PEGDME_CPE|S:SPC 70:30 w/w and Li|DOL:DME_LE|S:SPC 70:30 w/w cells at the constant current rate of C/10 (1C = 1675 mA $g_S^{-1}$) in the 1.7 – 2.8 V voltage range for the former and between 1.9 and 2.8 V for the latter at 50 °C through a MACCOR Series 4000 battery test system. EIS data were collected at upon galvanostatic cycling of a Li|PEGDME_CPE|S:SPC 70:30 w/w cell at the OCV condition and after 50 discharge/charge cycles. All CV, EIS and chronoamperometry data were collected using a VersaSTAT MC Princeton Applied Research (PAR, AMETEK) instrument.

*SEM and thermogravimetric analyses of the electrode/electrolyte assembly*

SEM images of a PEGDME_CPE membrane on a S:SPC 70:30 w/w electrode recovered after the CV measurement in the Li cell were collected using a Zeiss EVO 40 microscope with a $LaB_6$ thermionic source. TGA was carried out under a $N_2$ atmosphere and employing a heating rate of 5 °C min$^{-1}$ in the 25 – 800 °C temperature range, through a Mettler-Toledo TGA 2 instrument. Several samples were investigated by TGA: (i) PEGDME2000 powder, (ii) PEGDME_CPE membrane and (iii, iv) PEGDME_CPE on a S:SPC 70:30 w/w electrode in pristine condition and after CV. The samples were transferred very quickly from the glovebox to the SEM and TGA chambers for measurements, in order to avoid excessive exposure to the atmosphere and limit moisture absorption.

**Acknowledgements**

This work has received funding from the European Union's Horizon 2020 research and innovation programme Graphene Flagship under grant agreement No 881603, and the grant "Fondo per l'Incentivazione alla Ricerca (FIR) 2020", University of Ferrara. The authors acknowledge the project "Accordo di Collaborazione Quadro 2015" between University of Ferrara (Department of Chemical and Pharmaceutical Sciences) and Sapienza University of Rome (Department of Chemistry).

**List of tables**

**Table 1.** PEGDME_CPE bulk resistance ($R_e$) and S:SPC 70:30 w/w/PEGDME_CPE interphase resistances ($R_1$, $R_2$) obtained by nonlinear least squares (NLLS) analysis of the EIS data via the Boukamp software.[66,67] The EIS data have been collected at various temperatures during CV measurements on the Li|PEGDME_CPE|S:SPC 70:30 w/w cell. See the Experimental section for sample acronyms and Fig. 2 for relevant voltammetry profiles and Nyquist plots.

**Table 2.** S:SPC 70:30 w/w/PEGDME_CPE interphase resistance ($R_1$) obtained by nonlinear least squares (NLLS) analysis of the EIS data via the Boukamp software.[66,67] The EIS data have been collected at 50 °C during a galvanostatic cycling measurement on the Li|PEGDME_CPE|S:SPC 70:30 w/w cell. See the Experimental section of the manuscript for sample acronyms and Figure 4 (c, d) for relevant Nyquist plots.

**List of figures**

**Figure 1. (a)** Lithium transference number ($t^+$) at the selected operating temperature and **(b)** ionic conductivity at 50 °C of solid PEO_PE,[41] PEGDME_CPE, and DOL:DME_LE (see Experimental section for details on sample composition); **(c)** TGA and **(d)** DTG curves of the PEGDME_CPE and the PEGDME2000 powder recorded in a temperature range between 25 °C and 800 °C at a heating rate of 5 °C min$^{-1}$. See the Experimental section for sample acronyms.

**Figure 2. (a, c, e)** CV curves and **(b, d, f)** EIS Nyquist plots of the Li|PEGDME_CPE|S:SPC 70:30 w/w cell at various temperatures, that is, **(a, b)** 50 °C, **(c, d)** 60 °C and **(e, f)** 70 °C. CV performed between 1.8 and 2.8 V *vs.* Li$^+$/Li at 0.1 mV s$^{-1}$; EIS carried out at the open circuit voltage (OCV) condition of the cell as well as upon the voltammetry cycles, by applying an alternate voltage signal of 10 mV within the 500 kHz – 100 mHz frequency range. See the Experimental section for sample acronyms and Table 1 for relevant parameters extracted by analysis of the EIS data.

**Figure 3. (a, b)** Photographic images and **(c)** TGA curves with **(d)** corresponding DTG profiles of a PEGDME_CPE membrane on a S:SPC 70:30 w/w electrode (i.e., the PEGDME_CPE|S:SPC 70:30 w/w electrode assembly) **(a)** in the pristine condition (before assembling the cell) and **(b)** after CV in the lithium cell at various temperatures (see Fig. 2); TGA carried out in the 25 – 800 °C temperature range by employing a heating rate of 5 °C min$^{-1}$. **(e)** SEM image (left-hand side) showing the PEGDME_CPE|S:SPC 70:30 w/w electrode assembly after CV in the lithium cell at various temperatures (see Fig. 2), with related graphic scheme (right-hand side); SEM inset highlights the presence of the electrode layers, i.e., (i) the PEGDME_CPE membrane (between red lines) and (ii) the carbon-sulfur electrode film (between bottom-red line and cyan line).

**Figure 4.** Electrochemical performance of the Li|PEGDME_CPE|S:SPC 70:30 w/w cell at 50 °C in terms of **(a)** voltage profiles and **(b)** cycling trend (discharge capacity in left-hand side *y*-axis and coulombic efficiency in right-hand side *y*-axis) at the constant current rate of C/10 (1C = 1675 mA g$_S^{-1}$); voltage range: 1.7 – 2.8 V. **(c, d)** EIS measurements performed on a Li|PEGDME_CPE|S:SPC 70:30 w/w cell at 50 °C at various states, that is, **(c)** at the OCV condition and **(d)** after 50 discharge/charge cycles performed at C/10 (1C = 1675 mA g$_S^{-1}$) between 1.7 and 2.8 V. EIS frequency range: 500 kHz – 100 mHz; alternate voltage signal: 10 mV. **(e)** Photographic image of a PEGDME_CPE membrane on a S:SPC 70:30 w/w electrode recovered after the cycling test.

| Temperature (°C) | Cell condition | Circuit | $R_e$ [Ω] | $R_1$ [Ω] | $R_2$ [Ω] | $R_1 + R_2$ [Ω] | $\chi^2$ |
|---|---|---|---|---|---|---|---|
| 50 °C | OCV | $R_e(R_1Q_1)Q_w$ | 272 ± 1 | 83 ± 2 | / | 83 ± 2 | 5×10⁻⁶ |
|  | 1 CV cycle | $R_e(R_1Q_1)Q_w$ | 298 ± 2 | 44 ± 2 | / | 44 ± 2 | 4×10⁻⁶ |
|  | 3 CV cycles | $R_e(R_1Q_1)Q_w$ | 304 ± 3 | 42 ± 3 | / | 42 ± 3 | 3×10⁻⁶ |
| 60 °C | OCV | $R_e(R_1Q_1)(R_2Q_2)Q_w$ | 150 ± 2 | 36 ± 2 | 3.4 ± 0.5 | 39 ± 2 | 2×10⁻⁶ |
|  | 3 CV cycles | $R_e(R_1Q_1)(R_2Q_2)Q_w$ | 151 ± 2 | 15 ± 3 | 2.3 ± 0.9 | 17 ± 3 | 1×10⁻⁵ |
| 70 °C | OCV | $R_e(R_1Q_1)(R_2Q_2)Q_w$ | 102 ± 1 | 9 ± 1 | 3.9 ± 0.4 | 13 ± 1 | 8×10⁻⁷ |
|  | 3 CV cycles | $R_e(R_1Q_1)(R_2Q_2)Q_w$ | 90 ± 4 | 11 ± 4 | 3 ± 1 | 14 ± 4 | 9×10⁻⁶ |

**Table 1**

| Cell condition | Circuit | $R_1$ [Ω] | $\chi^2$ |
|---|---|---|---|
| OCV | $R_e(R_1Q_1)Q_w$ | 179 ± 10 | $8\times10^{-5}$ |
| After 50 cycles | $R_e(R_1Q_1)Q_w$ | 108 ± 17 | $3\times10^{-5}$ |

**Table 2**

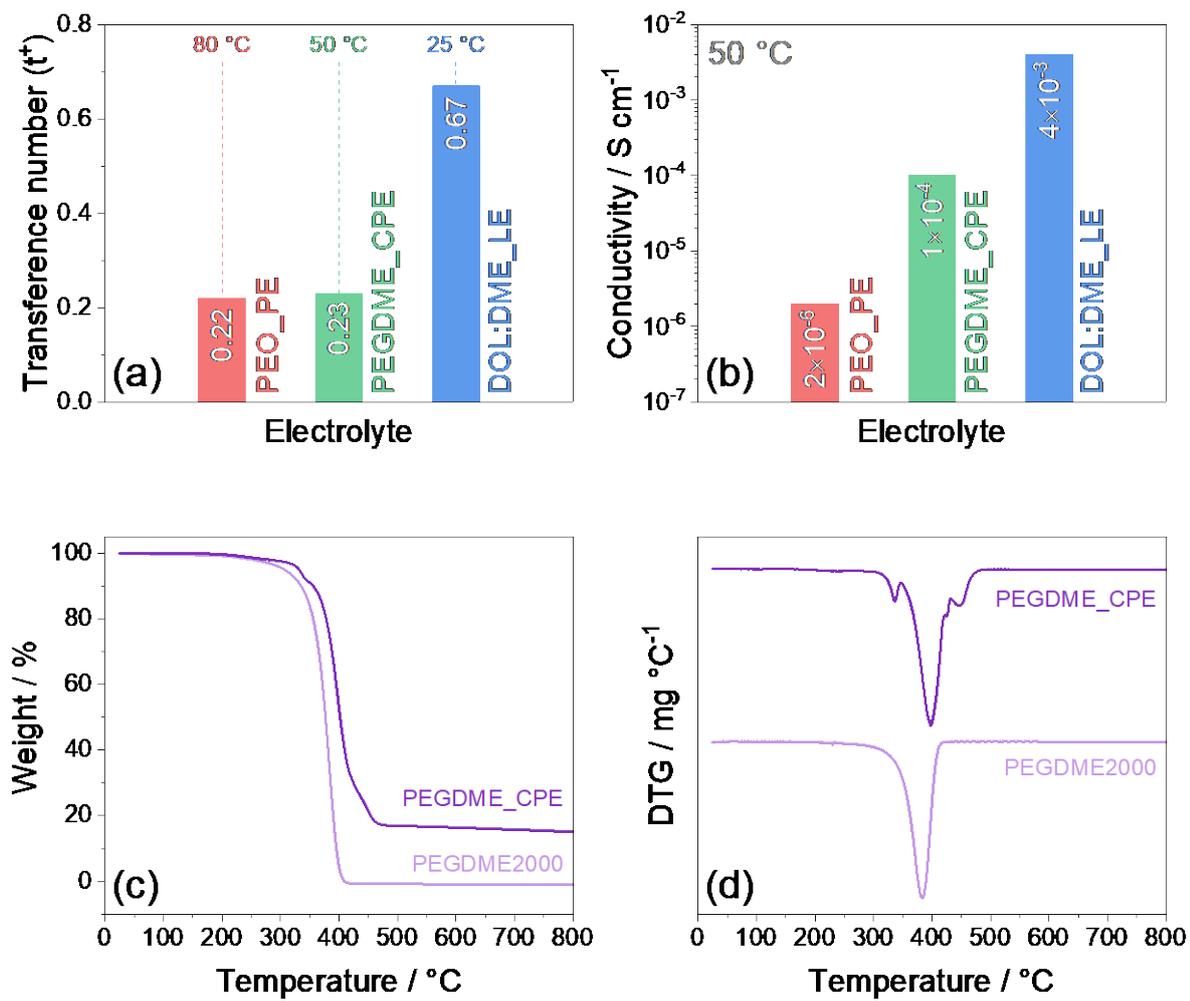

**Figure 1**

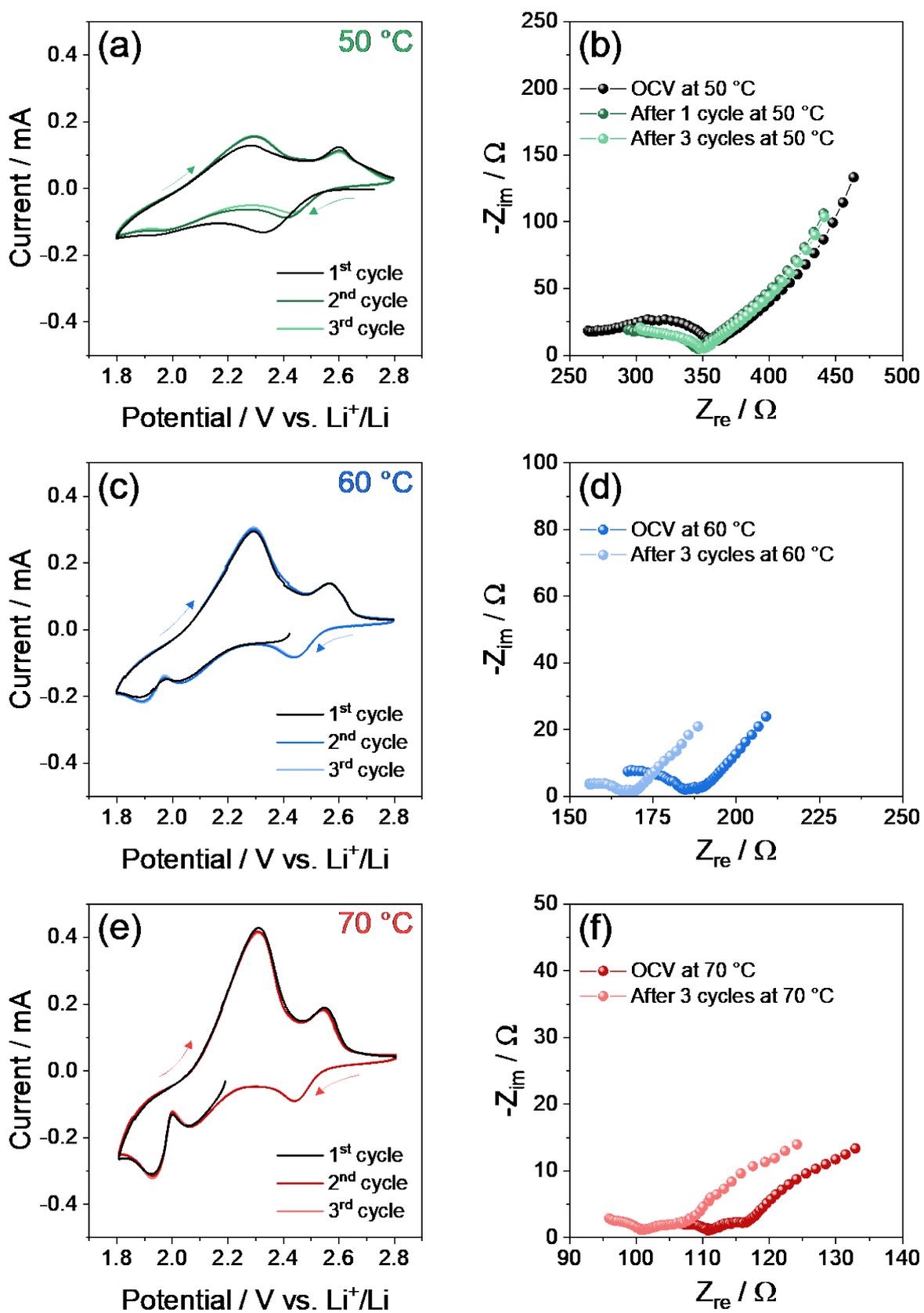

**Figure 2**

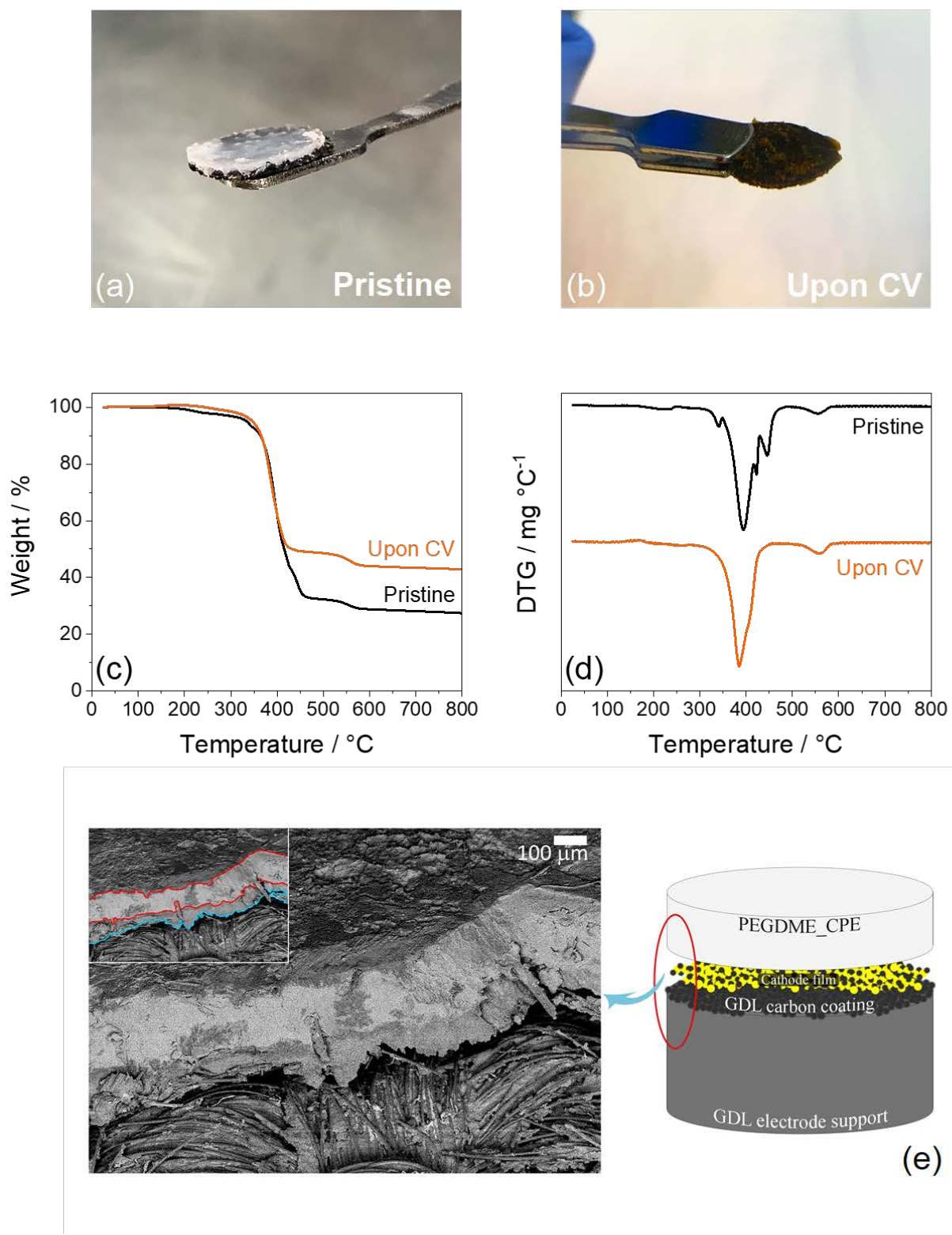

**Figure 3**

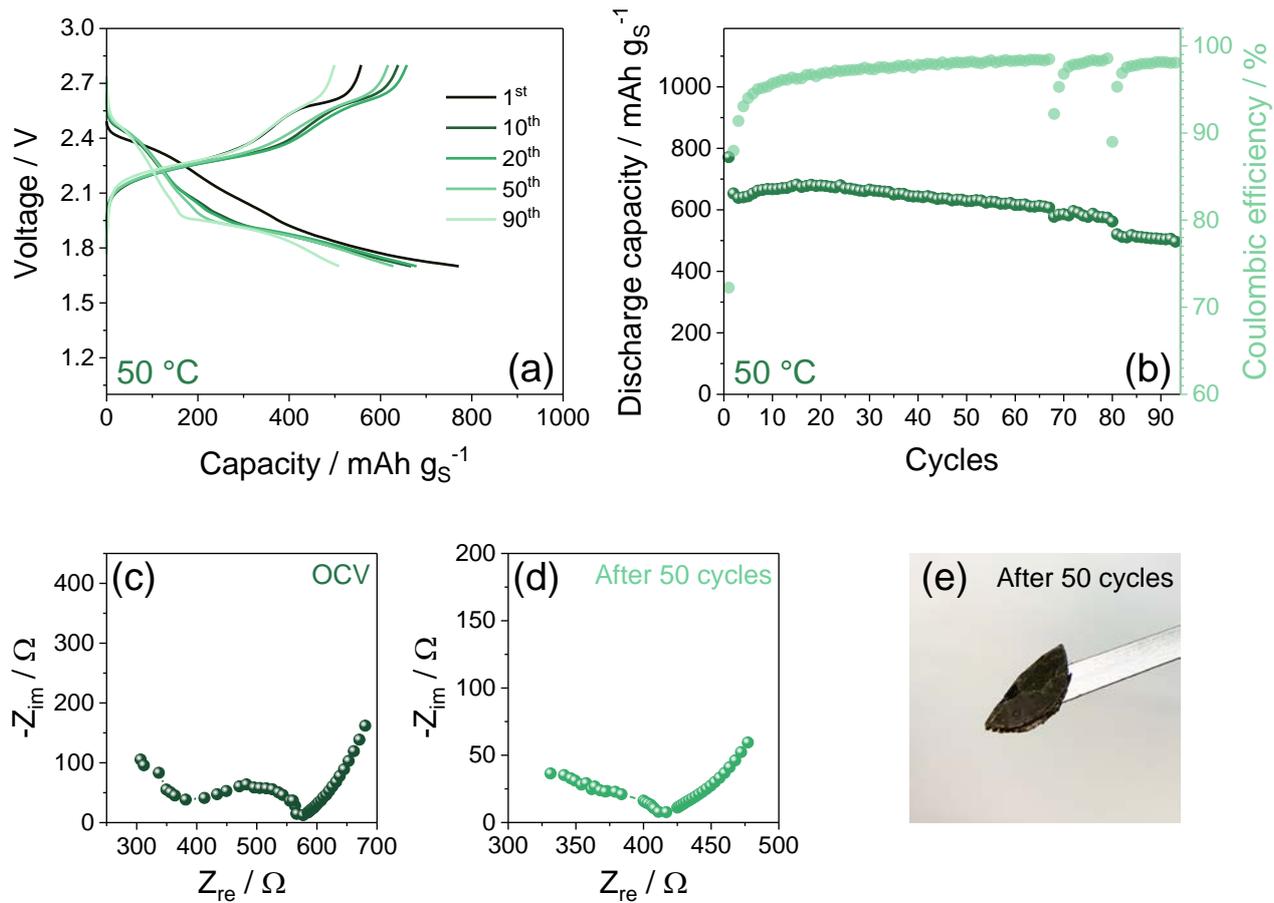

**Figure 4**

**Novel Lithium-Sulfur Polymer Battery Operating at Moderate Temperature**

Vittorio Marangon[a], Daniele Di Lecce[b], Luca Minnetti[b], Jusef Hassoun[a,b,c]*

[a] *Department of Chemical, Pharmaceutical and Agricultural Sciences, University of Ferrara, Via Fossato di Mortara 17, Ferrara, 44121, Italy*

[b] *Graphene Labs, Istituto Italiano di Tecnologia, via Morego 30, Genova, 16163, Italy*

[c] *National Interuniversity Consortium of Materials Science and Technology (INSTM), University of Ferrara Research Unit, Via Fossato di Mortara, 17, 44121, Ferrara, Italy.*

*Corresponding author. E-mail addresses: jusef.hassoun@unife.it, jusef.hassoun@iit.it.

**Supporting Information**

Figure S1 reports the results of chronoamperometry and electrochemical impedance spectroscopy (EIS) measurements performed to evaluate the Li$^+$ transference number at room temperature (t$^+$, Fig. S1a) and the ionic conductivity at 50 °C (Fig. S1b) of a DOL:DME-based liquid electrolyte (DOL:DME_LE, see experimental section of the manuscript for details on the electrolyte formulation). According to the method proposed by Evans *et al.*,[1] a voltage signal is applied on a symmetrical Li|DOL:DME_LE|Li cell (panel a), and impedance spectra are collected before and after polarization (Nyquist plots in panel a inset). Current and electrode/electrolyte interphase resistance before polarization (i$_0$ and R$_0$, respectively) and at the steady state (i$_{ss}$ and R$_{ss}$, respectively) are obtained, and the Bruce-Vincent-Evans equation[1] is applied (see equation (1) in Experimental section of the manuscript). These resistance values are obtained by nonlinear least squares (NLLS) analysis of the EIS data[2,3] and reported in Table S1 along with a calculated t$^+$ of 0.67, which is suitable for application in Li-S battery. However, we remark that the electrochemical determination of t$^+$ may be affected by experimental issues leading to an overestimation.[4]

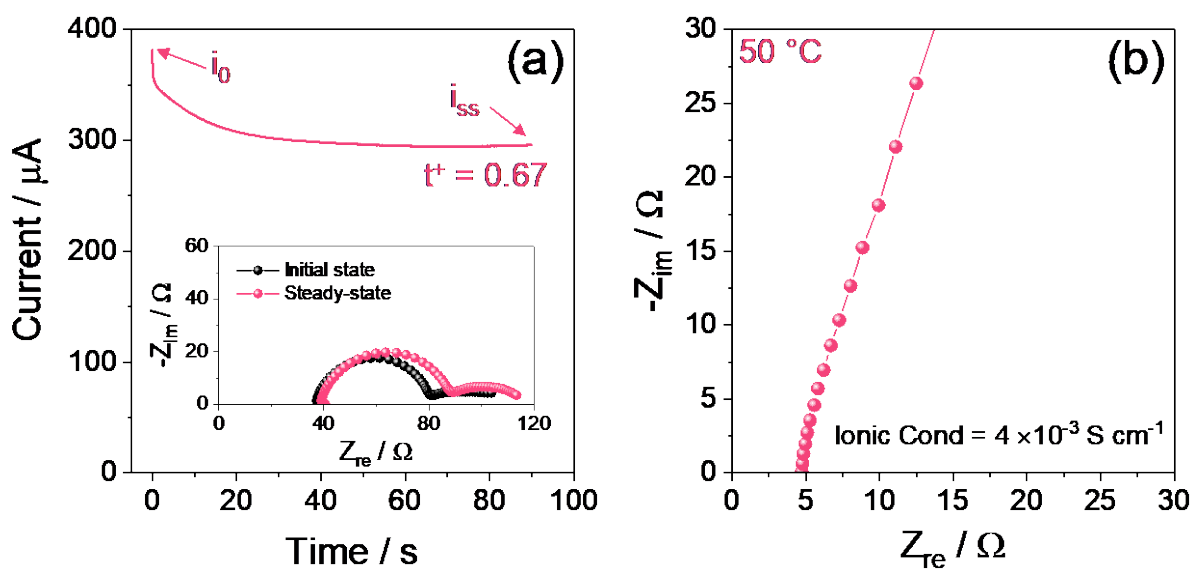

**Figure S1.** (a) Chronoamperometry profile and (inset) EIS Nyquist plot at room temperature of the symmetrical Li|DOL:DME_LE|Li cell before and after polarization; the Li$^+$ transference number of the electrolyte solution obtained by these data (t$^+$) is shown in the main panel.[1] Chronoamperometry voltage signal: 30 mV; EIS frequency range: 500 kHz – 100 mHz; alternate voltage signal amplitude: 10 mV. (b) EIS Nyquist plot at 50 °C of the symmetrical blocking cell using DOL:DME_LE; the ionic conductivity of the electrolyte solution obtained by these data is shown in the panel. EIS frequency range: 500 kHz – 100 Hz; alternate voltage signal: 10 mV.

The ionic conductivity of DOL:DME_LE at 50 °C is measured by EIS on a symmetrical blocking cell (Fig. S1b). According to EIS, the electrolyte has resistance of about 5 Ω, leading to an ionic conductivity of $4\times10^{-3}$ S cm$^{-1}$. A histogram plot of both t$^+$ and ionic conductivity of the DOL:DME_LE is reported in Figures 1a and 1b in the manuscript, respectively.

| Electrolyte | Initial current ($i_0$) [µA] | Steady-state current ($i_{ss}$) [µA] | Initial resistance ($R_0$) [Ω] | Steady-state resistance ($R_{ss}$) [Ω] | Transference number (t$^+$) |
|---|---|---|---|---|---|
| DOL:DME_LE | 381 | 296 | 44 | 49 | 0.67 |

**Table S1.** Parameters obtained by chronoamperometry and EIS data on the symmetrical Li|DOL:DME_LE|Li cell and used to calculate t$^+$ (see Fig. S1a).[1]

Figure S2 shows the Nyquist plot obtained by EIS measurement on a symmetrical blocking cell to evaluate the ionic conductivity of the PEGDME_CPE at 80 °C. At this temperature, the electrolyte exhibits a resistance of about 43 Ω and an associated ionic conductivity of $4 \times 10^{-4}$ S cm$^{-1}$. Thus, the measurement proves the applicability of PEGDME_CPE in the battery even at a temperature as high as 80 °C with suitable ionic conductivity value taking in consideration the solid configuration.

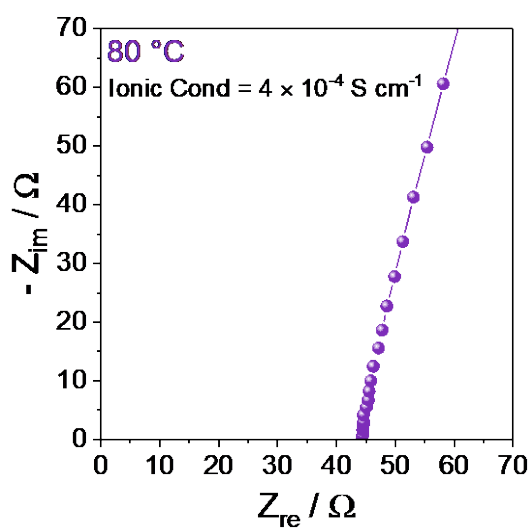

**Figure S2.** Nyquist plot recorded by EIS measurement at 80 °C on a symmetrical blocking cell employing the PEGDME_CPE; the corresponding ionic conductivity is displayed in the panel. EIS frequency range: 500 kHz – 100 Hz; alternate voltage signal: 10 mV.

The electrochemical process of the Li-S cell employing the PEGDME_CPE is investigated by cyclic voltammetry (CV) and EIS at various temperatures between 50 and 80 °C. The related voltammograms and recorded Nyquist plots are reported in Figure 2 of the manuscript (from 50 to 70 °C) and in Figure S3 (80 °C). Figure S3a reveals that the charge peaks merge into a double current wave with a shape similar to that observed for Li-S batteries exploiting liquid glyme-based electrolytes,[5] thereby suggesting a *liquid-like* response of the PEGDME_CPE in the cell at 80 °C. As discussed in the manuscript (see discussion of Fig. 2), this condition favors a fast $Li^+$ ions mobility through the electrolyte membrane and the electrode/electrolyte interphases and, thus, limits the polarization of the electrochemical process. Indeed, NLLS analysis of the EIS data collected at upon CV at 80 °C (Table S2 and Fig. S3b, respectively) evidences low resistance of both electrolyte ($R_e$, ~ 60 Ω) and electrode/electrolyte interphase ($R_1 + R_2$, between 7 and 12 Ω), as well as modifications occurring in the latter by cycling which are reflected as change in equivalent circuit.

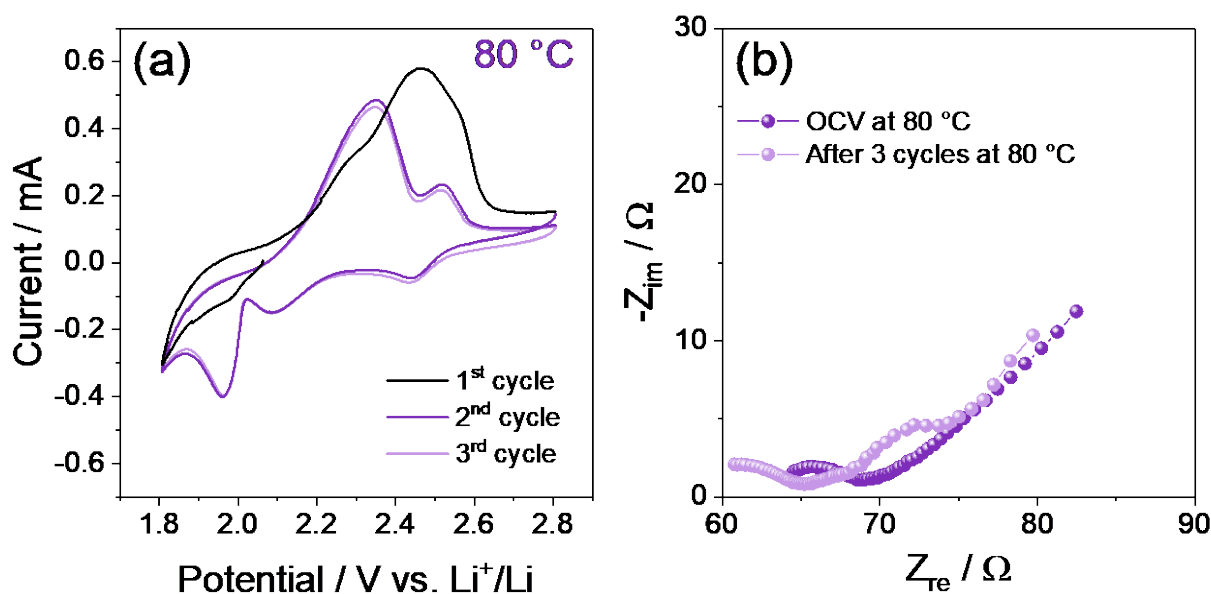

**Figure S3. (a)** CV profiles and **(b)** Nyquist plots of the Li|PEGDME_CPE|S:SPC 70:30 w/w cell at 80 °C. CV performed between 1.8 and 2.8 V *vs.* $Li^+$/Li at 0.1 mV $s^{-1}$; EIS carried out at the open circuit voltage (OCV) condition of the cell as well as upon the voltammetry cycles, by applying an alternate voltage signal of 10 mV within the 500 kHz – 100 mHz frequency range. See the Experimental section of the manuscript for sample acronyms and Table S2 for relevant parameters extracted by analysis of the EIS data.

| Temperature (°C) | Cell condition | Circuit | $R_e$ [Ω] | $R_1$ [Ω] | $R_2$ [Ω] | $R_1 + R_2$ [Ω] | $\chi^2$ |
|---|---|---|---|---|---|---|---|
| 80 °C | OCV | $R_e(R_1Q_1)Q_w$ | 61.0 ± 0.4 | 7.2 ± 0.4 | / | 7.2 ± 0.4 | 1×10⁻⁶ |
| | 3 CV cycles | $R_e(R_1Q_1)(R_2Q_2)(R_wQ_w)Q_w$ | 55 ± 1 | 10 ± 1 | 3 ± 2 | 12 ± 2 | 2×10⁻⁶ |

**Table S2.** PEGDME_CPE bulk resistance ($R_e$) and S:SPC 70:30 w/w/PEGDME_CPE interphase resistances ($R_1$, $R_2$) obtained by nonlinear least squares (NLLS) analysis of the EIS data via the Boukamp software.[2,3] The EIS data have been collected at 80 °C during a CV measurement on the Li|PEGDME_CPE|S:SPC 70:30 w/w cell. See the Experimental section of the manuscript for sample acronyms and Figure S3 for relevant voltammetry profiles and Nyquist plots.

Figure S4 shows the results of CV and EIS measurements at room temperature on a benchmark Li-S cell using standard DOL:DME_LE and S:SPC 70:30 w/w as the cathode. The CV profile of the first cycle (Fig. S4a) shows the typical multi-step, reversible process characterized by two reduction peaks at 2.3 and 2.0 V *vs.* Li$^+$/Li upon discharge and a broad double current wave between 2.3 and 2.5 V *vs.* Li$^+$/Li during charge.[6]

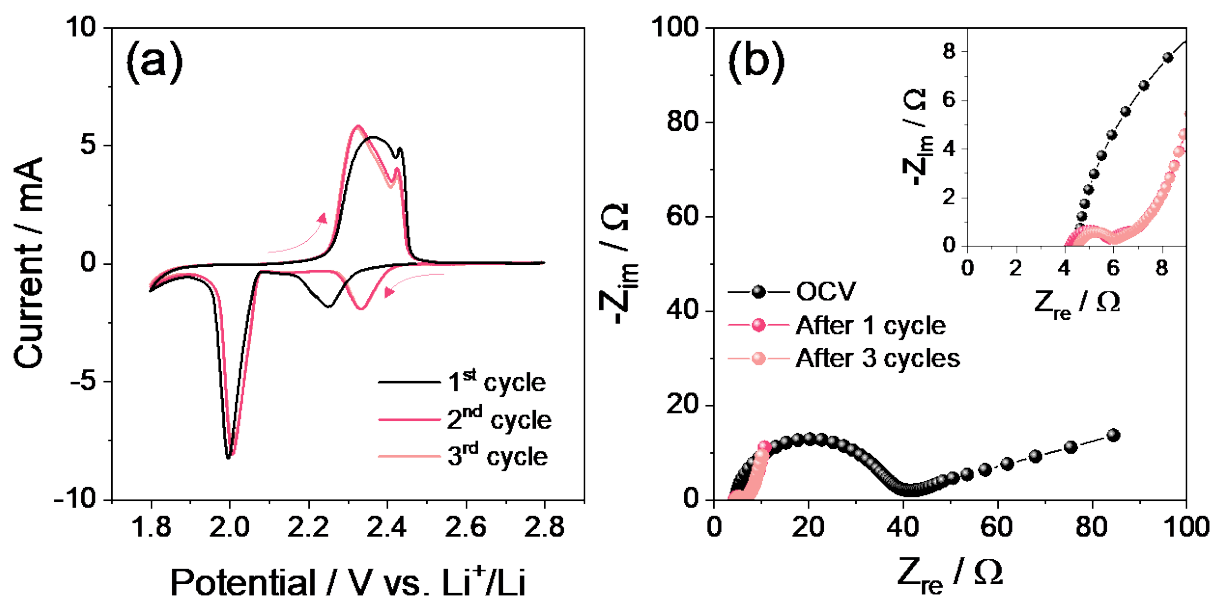

**Figure S4.** (a) CV profiles and (b) Nyquist plots of the Li|DOL:DME_LE|S:SPC 70:30 w/w cell at room temperature (25 °C). CV performed between 1.8 and 2.8 V *vs.* Li$^+$/Li at 0.1 mV s$^{-1}$; EIS carried out at the open circuit voltage (OCV) condition of the cell as well as upon the voltammetry cycles, by applying an alternate voltage signal of 10 mV within the 500 kHz – 100 mHz frequency range. See the Experimental section of the manuscript for sample acronyms and Table S3 for relevant parameters extracted by analysis of the EIS data.

A shift of the reduction peaks to higher potentials is observed upon cycling, and attributed to micro-structural rearrangements of the electrode/electrolyte interphase that enhance the conversion process.[6] Favorable modifications of the electrode/electrolyte interphase are also suggested by the EIS data (Fig. S4b and Table S3 showing the results of a NLLS analysis of the spectra),[2,3] which indicate a decrease of the electrode/electrolyte interphase resistance ($R_1 + R_2$) from 31 Ω at the OCV condition to values around 3 Ω after 3 CV cycles.

| Cell condition | Circuit | $R_e$ [Ω] | $R_1$ [Ω] | $R_2$ [Ω] | $R_1 + R_2$ [Ω] | $\chi^2$ |
|---|---|---|---|---|---|---|
| OCV | $R_e(R_1Q_1)Q_w$ | 3.2 ± 0.8 | 31 ± 1 | / | 31 ± 1 | 9×10$^{-4}$ |
| 1 CV cycle | $R_e(R_1Q_1)(R_2Q_2)Q_w$ | 4.3 ± 0.1 | 1.4 ± 0.1 | 1.3 ± 0.1 | 2.7 ± 0.2 | 2×10$^{-5}$ |
| 3 CV cycles | $R_e(R_1Q_1)(R_2Q_2)Q_w$ | 4.6 ± 0.1 | 1.3 ± 0.1 | 1.1 ± 0.2 | 2.4 ± 0.3 | 3×10$^{-6}$ |

**Table S3.** DOL:DME_LE bulk resistance ($R_e$) and S:SPC 70:30 w/w/DOL:DME_LE interphase resistances ($R_1$, $R_2$) obtained by nonlinear least squares (NLLS) analysis of the EIS data via the Boukamp software.[2,3] The EIS data have been collected at room temperature during a CV measurement on the Li|DOL:DME_LE|S:SPC 70:30 w/w. See the Experimental section of the manuscript for sample acronyms and Figure S4 for relevant voltammetry profiles and Nyquist plots.

Figure S5 and Table S4 illustrate the shifts in electrochemical potential of the processes in the Li|PEGDME_CPE|S:SPC 70:30 w/w cell with change in temperature, as observed by CV and described in the manuscript (discussion of Figure 2). Figure S5a identifies the charge and discharge signals in the voltammetry curve taken in consideration, while the corresponding trends of potential as a function of temperature are reported in Figure S5b-f. Notably, the 1$^{st}$ charge peak (Fig. S5b) gradually increases from 2.29 V *vs.* Li$^+$/Li at 50 °C to 2.35 *vs.* Li$^+$/Li at 80 °C, while the 2$^{nd}$ charge peak (Fig. S5c) decreases from 2.60 to 2.52 V *vs.* Li$^+$/Li. On the other hand, the 1$^{st}$ discharge signal (Fig. S5d) maintains a constant potential value around 2.44 V *vs.* Li$^+$/Li throughout the entire test, while the 2$^{nd}$ one (Fig. S5e) and the 3$^{rd}$ one (Fig. S5f) increase from 1.98 to 2.09 *vs.* Li$^+$/Li and from 1.89 to 1.96 *vs.* Li$^+$/Li, respectively. The potential shifts and the consequent decrease in polarization between the CV peaks enable us to quantify the enhancements of the electrode/electrolyte interphase by rise in the operating temperature described in the manuscript.

Figure S6 provides further evidence of the improvement of the Li|PEGDME_CPE|S:SPC 70:30 w/w system by cycling at increasing temperatures, that is, the trends of electrode/electrolyte interphase resistance (Fig. S6a) and bulk electrolyte resistance (Fig. S6b) as determined by the NLLS analyses[2,3] of EIS data collected after the third voltammetry cycle for each temperature (see Figure 2 and Table 1 in the manuscript, as well as Figure S3 and Table S2). Indeed, both resistance parameters decrease by rising the temperature, with values of 42 and 304 Ω at 50 °C for the former and the latter, dropping to 12 and 55 Ω at 80 °C, respectively.

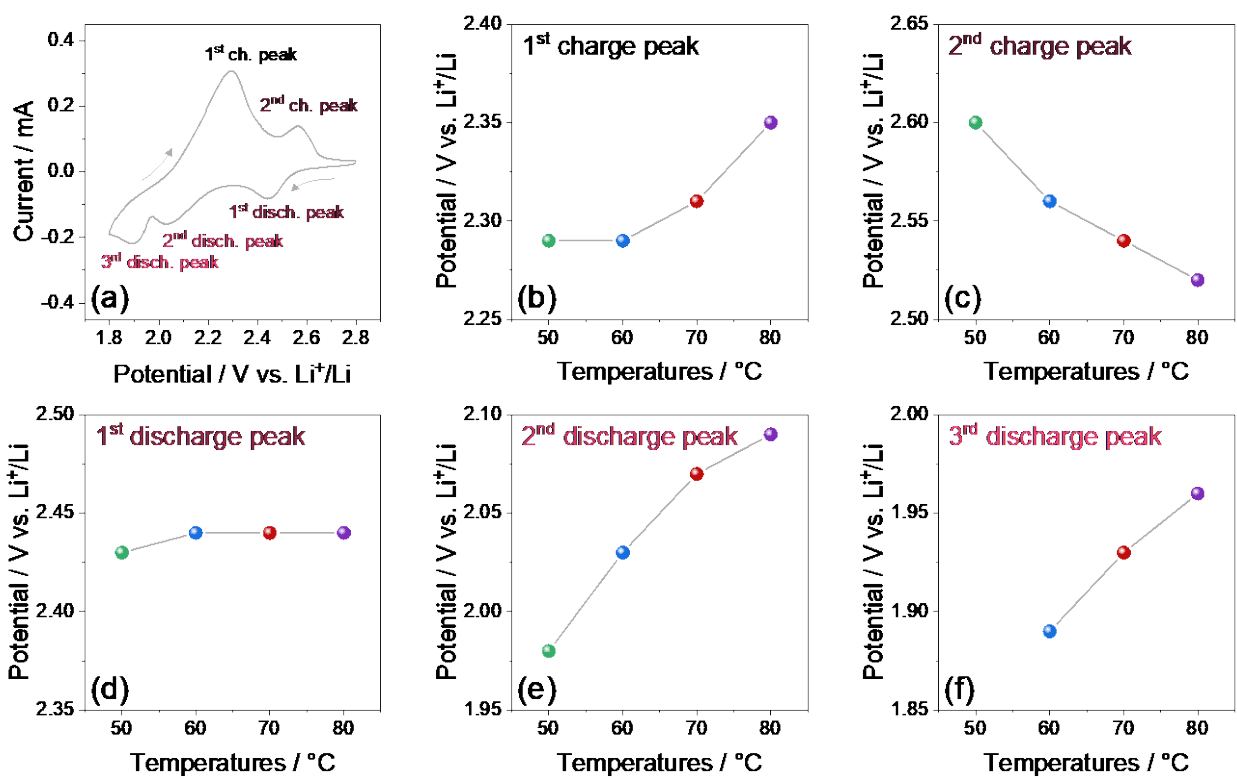

**Figure S5.** Trend of electrochemical potential of the Li|PEGDME_CPE|S:SPC 70:30 w/w cell at various temperatures from 50 to 80 °C. In detail: **(a)** CV profile of the cell with indication of the main current peaks, i.e., during charge **(b)** around 2.3 V *vs.* Li$^+$/Li (1st charge peak) and **(c)** between 2.5 and 2.6 V *vs.* Li$^+$/Li (2nd charge peak), and during discharge **(d)** at 2.45 V *vs.* Li$^+$/Li (1st discharge peak), **(e)** around 2.00 V *vs.* Li$^+$/Li (2nd discharge peak) and at about 1.90 V *vs.* Li$^+$/Li (3rd discharge peak). See the related CV profiles in Figure 2 of the manuscript (50 – 70 °C) and Figure S3 (80 °C), as well as the peak values in Table S4. All the potential values were measured during the 3rd cycle at each temperature.

| Temperature [°C] | 1st charge peak [V *vs.* Li$^+$/Li] | 2nd charge peak [V *vs.* Li$^+$/Li] | 1st discharge peak [V *vs.* Li$^+$/Li] | 2nd discharge peak [V *vs.* Li$^+$/Li] | 3rd discharge peak [V *vs.* Li$^+$/Li] |
|---|---|---|---|---|---|
| 50 | 2.29 | 2.60 | 2.43 | 1.98 | / |
| 60 | 2.29 | 2.56 | 2.44 | 2.03 | 1.89 |
| 70 | 2.31 | 2.54 | 2.44 | 2.07 | 1.93 |
| 80 | 2.35 | 2.52 | 2.44 | 2.09 | 1.96 |

**Table S4.** Electrochemical potential of the Li|PEGDME_CPE|S:SPC 70:30 w/w cell at various temperatures from 50 to 80 °C as measured by CV. See the related voltammetry curves in Figure 2 in the manuscript (50 – 70 °C) and Figure S3 (80 °C). All the potential values were measured during the 3$^{rd}$ cycle at each temperature.

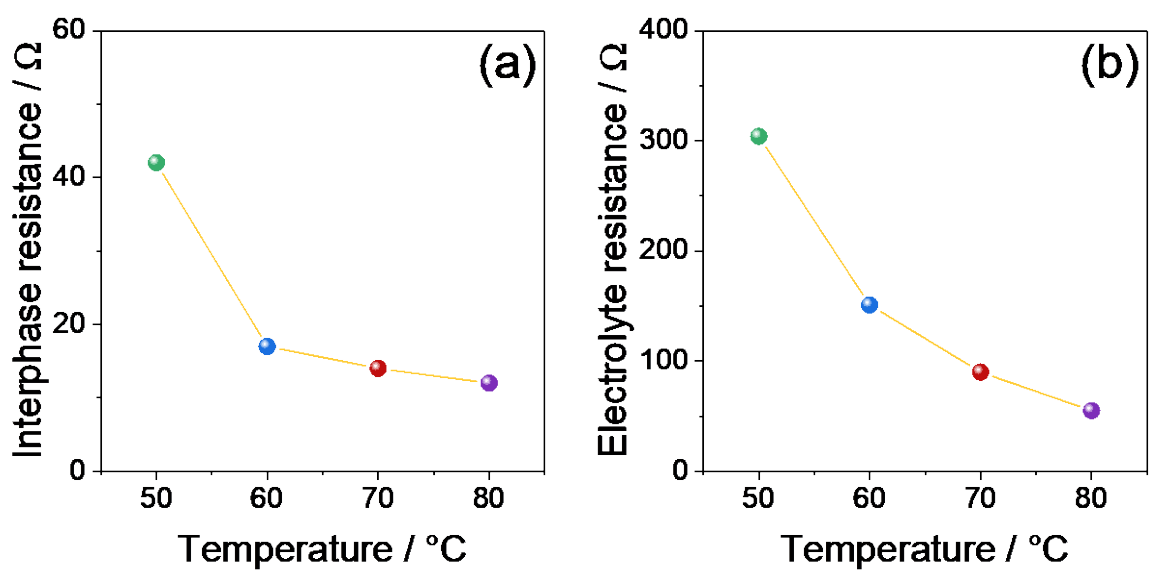

**Figure S6.** (a) Electrode/electrolyte interphase resistance and (b) electrolyte resistance in the Li|PEGDME_CPE|S:SPC 70:30 w/w cell at various temperatures from 50 to 80 °C, as determined by EIS. The impedance data were collected (see Figure 2 in the manuscript and Figure S3) and analyzed by the NLLS method.[2,3] These parameters are also reported in Table 1 in the manuscript and Table S2. All the resistance values were measured at each temperature after the 3$^{rd}$ cycle.

The applicability of the DOL:DME_LE at 50 °C was investigated through a galvanostatic test carried out at the current rate of C/10 (1C = 1675 mA g$_S^{-1}$) on a Li-S battery. Figure S7a displays the related voltage profile, and evidences the irreversible reduction of LiNO$_3$ at 1.9 V promoted by high temperature. This process is likely associated to the formation of a thick electrode/electrolyte interphase which leads to low delivered capacity of 900 mAh g$_S^{-1}$ in the cycling trend in Figure S7b. These data suggest the limited applicability of DOL:DME mixtures as electrolyte solvents at relatively high temperatures.

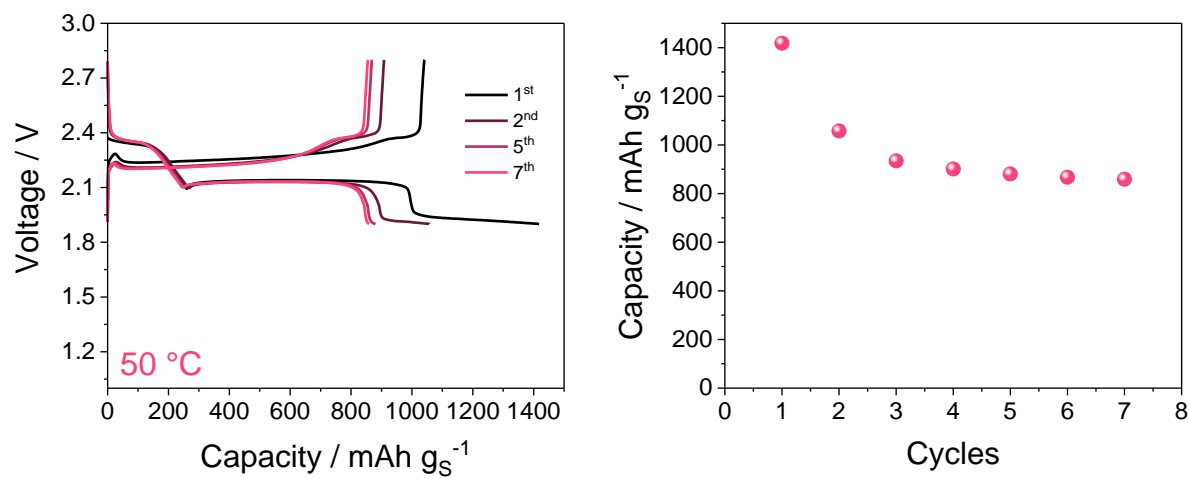

**Figure S7. (a)** Voltage profiles and **(b)** cycling trend of a Li|DOL:DME_LE|S:SPC 70:30 w/w cell galvanostatically cycled at the current rate of C/10 (1C = 1675 mA $g_S^{-1}$) between 1.9 and 2.8 V at 50 °C.